\documentclass[fleqn,10pt]{wlscirep}
\usepackage[utf8]{inputenc}
\usepackage[T1]{fontenc}
\title{Elevating Intrusion Detection and Security Fortification in Intelligent Networks through Cutting-Edge Machine Learning Paradigms}

\author[1,+]{Md Minhazul Islam Munna}
\author[1,+]{Md Mahbubur Rahman}
\author[2,3]{Jaroslav Frnda}
\author[4,*]{Muhammad Shahid Anwar}
\author[5,*]{Alpamis Kutlimuratov}
\affil[1]{Department of Computer Science and Technology, Beijing Institute of Technology, 5 Zhongguancun South Street, Haidian District, Beijing, 100081, China.}
\affil[2]{Department of Quantitative Methods and Economic Informatics, Faculty of Operation and Economics of Transport and Communication, University of Zilina, 01026 Zilina, Slovakia.}
\affil[3]{Department of Telecommunications, Faculty of Electrical Engineering and Computer Science, VSB-Technical University of Ostrava, 70800 Ostrava, Czech Republic.}
\affil[4]{Department of AI and Software, Gachon University, Seongnam-si, 13120, South Korea.}
\affil[5]{Department of Applied Informatics, Kimyo International University in Tashkent, Uzbekistan.}

\affil[*]{shahidanwar786@gachon.ac.kr}

\affil[+]{these authors contributed equally to this work}


\begin{abstract}
The proliferation of IoT devices and their reliance on Wi-Fi networks have introduced significant security vulnerabilities, particularly the KRACK and Kr00k attacks, which exploit weaknesses in WPA2 encryption to intercept and manipulate sensitive data. Traditional IDS using classifiers face challenges such as model overfitting, incomplete feature extraction, and high false positive rates, limiting their effectiveness in real-world deployments. To address these challenges, this study proposes a robust multiclass machine learning based intrusion detection framework. The methodology integrates advanced feature selection techniques to identify critical attributes, mitigating redundancy and enhancing detection accuracy. Two distinct ML architectures are implemented: a baseline classifier pipeline and a stacked ensemble model combining noise injection, Principal Component Analysis (PCA), and meta learning to improve generalization and reduce false positives. Evaluated on the AWID3 data set, the proposed ensemble architecture achieves superior performance, with an accuracy of 98\%, precision of 98\%, recall of 98\%, and a false positive rate of just 2\%, outperforming existing state-of-the-art methods. This work demonstrates the efficacy of combining preprocessing strategies with ensemble learning to fortify network security against sophisticated Wi-Fi attacks, offering a scalable and reliable solution for IoT environments. Future directions include real-time deployment and adversarial resilience testing to further enhance the model's adaptability.
\end{abstract}
\begin{document}

\flushbottom
\maketitle
%
%
\thispagestyle{empty}


\section*{Introduction}
\label{intro}

Machine Learning (ML) is pivotal in revolutionizing intelligent programmable networks, enabling them to meet the escalating
demands of modern, complex environments such as 5G and beyond. These smart networks enable dynamic traffic routing, adaptive bandwidth allocation, and real-time anomaly detection, making them pivotal for applications spanning smart homes, industrial automation, healthcare systems, and autonomous vehicles \cite{nazir2024enhancing, khan2024energy}. Among the core technologies enabling these ecosystems is Wi-Fi, which facilitates wireless communication for billions of devices. However, this convenience comes at the cost of growing vulnerability. Wi-Fi protocols, especially WPA2, have been repeatedly shown to suffer from critical flaws. In particular, key reinstallation attacks (KRACK) and KR00K, exploit weaknesses in WPA2’s four-way handshake and chipset-level key handling, respectively, allowing adversaries to intercept, decrypt, or manipulate sensitive network traffic \cite{salah2023, alraih2022}. These attacks are particularly concerning in IoT environments, where devices often lack frequent firmware updates, making them persistently vulnerable \cite{nandhini2024}.

Over the years, numerous machine learning (ML)-based IDS have been developed to detect and mitigate such network threats. Classical models, such as Decision Trees, Support Vector Machines, and k-Nearest Neighbors, have been used in combination with packet-level feature extraction, overfitting issues, and flow-based heuristics to detect anomalous behavior \cite{ahn2021, mohan2022}. However, when it comes to Wi-Fi-specific threats like KRACK and KR00K, traditional IDSs face several limitations. First, these attacks often operate at the link layer and do not induce large deviations in network traffic patterns, making them difficult to detect using standard anomaly-based techniques \cite{Chatzoglou2021}. Second, existing models are highly sensitive to feature noise, suffer from overfitting on imbalanced data sets, and are not optimized for multiclass classification \cite{salah2023}. Third, many detection systems generate high false positive rates (FPR), which renders them impractical for real-time environments where false alarms may overwhelm network administrators \cite{alperin2021}. Lastly, the feature extraction methods employed in previous works often lack robustness in generalizing across varied device types, operating systems, and chipsets, factors that are crucial for widespread adoption in real-world settings \cite{kolias2015}.

Given these shortcomings, there is an urgent need for a more comprehensive and noise-resilient intrusion detection framework that can handle the subtlety of WPA2-based attacks while minimizing false alarms. Our research is motivated by this gap. We argue that detecting KRACK and KR00K attacks requires not just powerful classifiers but also a robust data preprocessing pipeline that can denoise input features, reduce dimensionality, and enhance generalization \cite{Chatzoglou2021}. Furthermore, ensemble learning, particularly stacking-based approaches, offers a promising route to combine the strengths of multiple classifiers while minimizing individual weaknesses \cite{nandanwar2024}. When coupled with advanced techniques such as Principal Component Analysis (PCA) for feature compression and Gaussian noise injection for regularization, these ensemble models can significantly improve detection performance on noisy, high-dimensional, and imbalanced data sets such as AWID3 \cite{salah2023, nandhini2024}.

To this end, we propose a two-tier machine learning pipeline for wireless intrusion detection. The first tier (Pipeline 1) trains and evaluates individual classifiers, including Support Vector Machines (SVM), Random Forests (RF), XGBoost, k-NN, and Multi-Layer Perceptrons (MLP), under controlled preprocessing settings. The second tier (Pipeline 2) extends this by injecting Gaussian noise into input features, applying PCA to preserve 90\% of data variance while reducing redundancy, and combining probabilistic outputs from base learners into a meta-feature vector. This vector is then passed to a meta-classifier, implemented using XGBoost, to make the final prediction. By adopting this ensemble architecture, the system benefits from diversified decision boundaries and improved stability \cite{abdulganiyu2025}. Additionally, we apply techniques such as Variance Inflation Factor (VIF) analysis to eliminate multicollinearity and Synthetic Minority Oversampling Technique (SMOTE) to address class imbalance \cite{adeyiola2023, salah2023}. These enhancements ensure that the model is not only accurate but also generalizable and scalable for deployment in IoT-centric networks.

\noindent \textbf{The key contributions of this study are as follows:}

\begin{itemize}
    \item We introduce a novel ensemble-based IDS that integrates noise augmentation, PCA, and stacked meta-learning to enhance detection of KRACK and Kr00k attacks in Wi-Fi traffic.
    \item We apply Variance Inflation Factor (VIF) analysis to eliminate multicollinearity and Synthetic Minority Oversampling Technique (SMOTE) to balance class distributions, significantly improving model stability.
    \item Our stacked ensemble pipeline (Pipeline 2) achieves 98\% accuracy with a reduced false positive rate of 2\%, outperforming baseline models with FPRs between 4–7\%, thereby improving real-world deployment feasibility in security-critical environments.
\end{itemize}

\section*{Related Works}
\label{sec:Related Works}
This malfunction has hit Wi-Fi's most prominent vulnerabilities, Krack (Key Reinstallation Attack) and Kr00k. The Krack attack exploits a vulnerability in the four-way handshake process established by the WPA2 protocol that allows secure communication between devices. The attacker then attacks this flaw - reinstalls the decryption key - and is able to intercept sensitive information\cite{ahn2021}. The Kr00k, on the other hand, can be understood as a vulnerability that affects Wi-Fi chips from Broadcom and Cypress models; all these chips go through the process of resetting the encryption key into a zero momentarily during a disassociation-an act that makes it possible for an attacker to capture and decrypt information off the data packets\cite{alraih2022}.
These weaknesses have created serious cracks in wireless network security and indicated the need for sophisticated IDS. Such attacks cannot usually be detected by using conventional security strategies such as firewalls and encryption, compelling researchers to look at more advanced techniques for improving detection accuracy and damage mitigation, such as ML \cite{mohan2022}. ML is one of the best available approaches for intrusion detection because it can be used in analyzing huge amounts of data and finding patterns that lead to the discovery of possible security threats. Quite a lot of scientific work has been wasted in using ML techniques to build WIDS, using databases such as AWID which contains data from the real world on network traffic - even with examples of various attacks \cite{ahn2021, prabha2022}.

For instance, a ML-based approach to detect Krack and Kr00k attacks using the AWID3 data set, which includes detailed traces of wireless network activity\cite{borgaonkar2021}. Their study highlights the efficacy of ML models in identifying these vulnerabilities, achieving a detection accuracy of 99\% for Krack attacks and 96.7\% for Kr00k attacks\cite{park2021}. The authors used ensemble classifiers and neural networks, demonstrating the potential of ML in enhancing wireless network security\cite{salah2023}. The performance of a deep learning model in building an IDS that identifies network intrusion types including attacks based on Wi-Fi was understood to be similar to a study. This model registers a 98 percent accuracy rating, demonstrating deep learning effectiveness in the identification of sophisticated patterns in attack\cite{salah2023}. Research in this regard shows that ML techniques, especially in deep learning, have significant improvement in the efficiency and accuracy of intrusion detection mechanisms\cite{gonzalez2021, mohan2022}.

Dealing with the high dimensionality of noise and the high measurement dimensions of several traffic data is the primary challenge for ML in intrusion detection. Feature selection techniques will enhance model performance by reducing irrelevant or redundant features. Preprocessing of the AWID data set that emphasizes the usage of the study beforehand into ML algorithms is one. ANOVA based feature selection improved its detection capability from 254 features down to 15. This preprocessing stage dramatically enhanced the model output and stressed the importance of feature selection in ML based IDS \cite{salah2023}. The issue of high dimensional data in their wireless IDS research. Thus, applying feature selection techniques improved the accuracy of the model to 99.67\%. This approach not only reduced computational complexity but also minimized overfitting, which is common among ML models using very large data sets. The evolution of such sophisticated attacks now on the wireless networks, namely Krack and the more recent Kr00k, has called for the design of new advanced IDS in them. In such a scenario, ML would realize a success story being regarded as the pillar on which high apparent accuracy in detection and mitigatemeasures rely.Within this context, supervised learning algorithms such as Decision Trees, Random Forests, and Neural Networks have widely found applications in anomaly detection\cite{garika2023iot}. With the inclusion of ML, WIDS perform even better in discovering last-minute or zero-day-attacks. In a proposed design of a hybrid WIDS, the system has a rule-based detection approach as well as an ML-based one. The system was able to recognize up to 98.57\% WLAN anomalies. In their approaches, the ML models were trained with the normal and attack scenarios of actual Wi-Fi traffic in real-time which was developed from AWID3, a very popular wi-fi based data set. Differentiating the attacks such as flooding, injection, and impersonation, the data sets helped in characterizing the attack models\cite{kumar2022}.

The IoT has ripped apart the previous communication records; these things enhance the ways devices can connect in terms of convenience and efficiency \cite{garika2023iot}. But there is a cloud of rapid proliferation of IoT devices creating new security risks. Often the IoT devices depend on Wi-Fi networks, so it becomes vulnerable to various attacks. One among those is Kr00k, a very high-ranked threat towards Wi-Fi security. The Kr00k attack exploits vulnerabilities found in Wi-Fi chips, especially those designed to use WPA2 encryption; it allows attackers to decrypt the data transmitted over the network. The current review discusses the existing studies pertaining to Kr00k attacks and highlights the application of ML in detecting and mitigating such attacks by focusing on Binary Grasshopper Optimization Algorithm and Long Short-Term Memory models proposed within recent studies\cite{nandhini2024}. The infrastructure of most IoT devices is based on Wi-Fi technology, which connects between devices and centralized systems through wireless technology. Despite all these, security problems have crept in with the blue wave of Wi-Fi networks available at home and office environments. Here comes the most common attack, the Kr00k attack, which is due to the vulnerabilities in Wi-Fi chips to decrypt any data. This affects billions of devices, right from the smartphones to IoT devices, and illegally penetrates through user information. Recently, traditional security measures, such as firewalls and encryption, are no longer enough to protect against the constant increase of changing dynamics in the network threat landscape. Thus, researchers have turned to applications of ML  and its techniques to improve IDS. In fact, many ML models, including Convolutional Neural Networks and LSTMs, have shown considerable promise in finding patterns in network traffic that point to security violations. These models also have many different limitations, including the problem of overfitting and missing some critical feature during feature extraction. Different ML algorithms applies to AWID, a data set for various Wi-Fi scenarios \cite{kolias2015}. They had selected the 20 features manually and passed them to 8 classifiers to identify Wi-Fi attacks\cite{zhao2023}. The accuracy of their models lay between 89\% and 96\%, but the selection of features was time-consuming and tedious. Stacked autoencoders (sae)\% were used in the same way for invasion detection on awid data set and achieved below expectation precision and recall \cite{thing2017}.

In another piece of research which employs detection modes with repositories of very high numbers within the seven-layer deep neural network (DNN) for malicious detection issues while the model tends to have higher false positive rates\cite{nandhini2024}, it has been proposed to focus on MCA-LSTM along the Temporal Co-relations amid intrusion data but failed to include space references as required, thus resulting in less accuracy\cite{agarwal2022, wang2018}. These examples reflect the paradox of forcing detection accuracy at a reasonably optimum rate of computation in network security models\cite{thing2017}. To address the problems of ML model feature extraction and overfitting, a new novel approach is developed in order to incorporate Binary Grasshopper Optimization Algorithm (BGOA) with LSTM models. BGOA is a very powerful and efficient optimization method inspired by the swarming behavior of grasshoppers. It can identify relevant features in a data set, which is critical in ML models regarding performance improvement as they enhance the identification of the most pertinent features contained in a data set. In the aspect of Kr00k attack detection, BGOA is guided toward selecting certain relevant network traffic features, which would have a big contribution toward accurate attack classification\cite{nandhini2024}.

The proposed approach improves network security much since it reduces the number of dimensions of the data set and thus improves model accuracy. Similarly, Nandhini et al. utilized the AWID3 data set with the huge number of 199,984 instances having 254 different features. BGOA had been applied to derive the most relevant features as inputs to an LSTM network for classification \cite{gonzalez2021}. The model managed to achieve 96\% accuracy at levels of precision and recall of 0.97 and 0.95 respectively, with an F1 score of 0.93, hence proving the strength of BGOA in optimizing features. LSTM networks can give an accurate interpretation of dynamic sequential data, which includes classifying network traffic and identifying patterns or clues that can be used to signal an intrusion. The proposed scheme makes use of LSTM networks for the classification of the Wi-Fi traffic as normal or malicious with respect to the Kr00k attack. The information-containing time dependence of network traffic by LSTM along with the best-relevant features selected by BGOA leads to a more accurate result of classification\cite{kumar2022}. The combination of CNN and LSTM models has proved successful in past efforts for use in attack detection\cite{kubota2020}. The main problem was that they were put together linearly and therefore limited the full use of temporal and spatial features, thus leading to a poorly performing model\cite{abdalgawad2021, kumar2022}. The combining of BGOA with LSTM would improve on selection of features and accuracy in classification. With the proliferation of more IoT devices, there is a pressing need for a strong network security solution. Kr00k attacks present a serious threat to Wi-Fi networks, and thus, developing ML models took a lot of time due to challenges faced during the selection of features and accuracy issues. BGOA is incorporated into feature selection with LSTM for attack classification. This holds great potential as the proposed method enhances the accuracy of the system in detection and also diminishes the computational complexity involved in processing large data sets. Future works should also focus on the optimization of these models for real-time applications with different network environments.

Wireless Local Area Networks (LANs) have become indispensable to today's modern communication infrastructure, forming the backbone for internet connectivity and data exchange. However, with increasing dependence on wireless networks inevitably comes concern and anxiety regarding security\cite{cermak2020}. One of the very recent and notable vulnerabilities that came up was a Kr00k attack, which exploits weaknesses found in Wi-Fi encryption protocols, particularly with WPA2 and WPA3. This review highlights some studies and methodologies revolving around detection and mitigation of Kr00k attacks, especially focusing on the combination of Channel Switch Announcement (CSA) and Kr00k attack methods\cite{nakajima2022}.
Kr00k, first revealed in 2020 by researchers at ESET, exploits a backdoor in the Wi-Fi chips used in billions of devices\cite{nakajima2022}. This vulnerability allows attackers to intercept and decrypt wireless communication simply by forcing the reset of the encryption key to zero during disassociation. Kr00k attacks WPA2-encrypted networks, which allow attackers to capture sensitive data like IP addresses or even entire packets of data by disassociating and reconnecting Wi-Fi devices\cite{kubota2020}.
Because this security flaw allows an attacker to trap leftover residual packets in the transmission buffer (Tx buffer) even when disconnected from the network, it becomes evil. As soon as the device reconnects, its contents become encrypted with a key value of 0, thus making it very easy to decrypt by attackers\cite{nakajima2022}. The vulnerability poses a significant threat; however, in reality, it is much more difficult to achieve Kr00k attacks because packets are sent instantaneously in most real-life scenarios\cite{kubota2019}.
Kr00k attacks work, but they usually boast poor success rates in practical setups. This is mostly because most packets are sent out within a very short period after entering the Tx buffer, so there is very little time available for hackers to capture them as they are\cite{konings2009}. In addition, once the users get disconnected, they may be able to notify others of the attack when they notice disconnections in the network during the attack attempts. This, thus, makes it difficult for the attacker to keep things prolonged and stealthy\cite{nakajima2022}.

There are some situations indicated by researchers-such as video streaming-in which there is a greater likelihood that packets will hold out longer in the Tx buffer, making Kr00k attacks plausible. Because streaming applications continuously transmit data, packets can build in the buffer for a little while; however, due to these limitations, the actual feasibility of a Kr00k attack is low without additional exploitation\cite{louca2021}. 
To address the drawbacks associated with the standard Kr00k attack, a new hybrid attack called Kr00k and CSA was proposed by researchers from Kobe University. Here, CSA refers to Channel Switch Announcement, a mechanism in a particular Wi-Fi network that notifies clients with information on future changes with respect to a particular channel. More importantly, there are indications that attackers send a corrupt CSA which tells the client to get disconnected from that network and switch to a non-existent channel. As a result, the device is forced to carry all the packets in the Tx buffer before disconnection and thus increases the chances of success for Kr00k\cite{nakajima2022}. The joint attack strategy has multiple benefits when compared to the traditional Kr00k attack. Firstly, attackers can exploit CSA, making clients buffer packets for longer durations so that interception becomes easier. Further, automatic reconnection of the client after receiving a modified CSA makes it possible to carry on with the attack without having the user alerted to any network disturbances. As such, the CSA-Kr00k attacks are very powerful since they can now be done for long durations without detection. Tests the researchers conducted showed how the CSA-Kr00k attack is viable in a real-world environment-theoretical scenarios involving all kinds of client devices, including Android and iOS, were involved. It could capture tons of data from the video streaming activity among users of the different devices. The data included sensitive information like source and destination IPs which would be good for hacking/dos purposes\cite{kubota2020}.

Interestingly, the researchers found that the CSA-Kr00k attack was more successful during live streaming sessions than during on-demand video streaming. This is because live streaming requires continuous data transmission, which increases the likelihood of packets being stored in the Tx buffer. On the other hand, on-demand video streaming transmits data in larger, less frequent bursts, reducing the probability of intercepting residual packets[21].The results also showed that different devices had varying levels of vulnerability to the attack. For example, certain devices like the Nexus 6P were more resilient due to their inability to process the CSA tampering. However, most other devices tested were vulnerable, highlighting the widespread risk posed by this attack method\cite{torres2021}.
As countermeasures against Kr00k and CSA-Kr00k attacks, a number of measures have been recommended. One of the most straightforward solutions would be to apply the patches made available by the manufacturers to address the vulnerability in affected devices; however, older devices and those on public Wi-Fi networks are not likely to receive such updates. In this case, users are advised to exercise necessary risks when connecting to unsecured networks and ensure their devices are using the latest encryption protocols\cite{ran2019}. In addition, enabling CSA or limiting the number of CSA signals that can be processed by a device can minimize the possibility of tampering. These measures would probably conflict with legitimate network operations and thus would not be practical in many situations. Ultimately, a mix of vendor updates, user awareness, and network configuration modifications is required for addressing completely the vulnerabilities introduced by Kr00k and CSA attacks\cite{duan2021}.
The Kr00k vulnerability, especially in combination with CSA tampering, creates a considerable risk to Wi-Fi security. The typical Kr00k attack has little real-world feasibility, but the CSA-Kr00k approach dramatically increases the chance of success and stealth. Continuous monitoring, as well as patch deployments, are significant requirements needed to mitigate these threats. Further research can work on more robust encryption and communication protocols that would counter such attacks from compromising wireless network security.
The surge in demand for wireless networks-wifi technology in particular-has generated a lot of security concerns. When there is communication with the outside world, wherein many IoT devices are connected, the security of Wi-Fi has become critical to the process. Many people use these wifi networks for daily communication. Security should be taken seriously.

Every organization has a robust IDS to help them detect the most unusual and unexpected attacks in their networks. This research literature review will analyze ongoing works on detection of intrusion into WiFi while considering activity anomaly detection analysis using online learning techniques with special emphasis to be laid on advances based on ML methods to improve network security. Therefore, it is physical and data link layers, where WiFi operates within OSI model convolutions, which expose it mostly to attacks because that exploit weakness of these link layers. The available security encryption methods to secure any network communications--like WPA2 and WPA3--are not readily proved to be effective in foiling attacks on the physical layer, where pathways for the new communication are always opened and closed. Hence, encryption should never be enough by itself to meet Confidentiality, Integrity, and Availability (CIA) requirements in wireless networks\cite{qin2018}. To mitigate the situation according to these needs, much research has been dedicated to designing various techniques for an IDS. It may be classified broadly into two types: Signature-Based IDS and Anomaly-Based IDS. The comparison of network traffic with signature databases can spot attacks in a signature-based IDS. It does have limitations such as the failure to identify new or modified attacks that differ from known signatures. In contrast, Anomaly Based IDS detects unusual activities that reveal deviation from the normal behavior of the network, thus making it more susceptible to identifying novel or zero-day attacks\cite{torres2021}. It is indeed true that ML  can and is being used in making personal IDS: to detect patterns in attacks and to classify them with remarkable accuracy. Many of the studies have been related to applying ML for anomaly detection in WiFi using immense data sets such as the Aegean WiFi Intrusion data set (AWID 2 and AWID 3). These data sets classify various types of attack traffic in normal operation and offer an immensely broad basis for a model to be trained on. 
Selection methods based on employed features are assessed with regard to a better detection tendency while reducing processing time \cite{abdulhammed2018}: Extra Trees ensemble method reduced that 20 features important from the AWID feature set, which are later injected into several different classifiers, such as a random forest and bagging. This study demonstrated that a feature-reduction approach could lead to better accuracy and speed for detection \cite{ran2019}.   
Analogously, Correlation based feature selection compresses the feature set of 156 down to 18 attributes. The research conducted an analysis of classifier accuracy between random forests and XGBoost. The major conclusion drawn from the research was that random forests outperformed the rest of the classifiers in accuracy terms. This is indicative of the role of feature selection in improving performance in ML models meant for intrusion detection\cite{vaca2018}. 

In neural networks, Some authors have investigated the application of deep learning models in WiFi intrusion detection\cite{duan2021}. The research suggests that neural networks, as highly accurate, tend to consume an enormous amount of computational resource and need a significantly higher number of features, making them rather infeasible for real-time detection systems\cite{feng2018}. The increasing vulnerability of Wi-Fi networks, particularly due to attacks like Krack and Kr00k, has triggered significant research into machine learning (ML)-based Wireless Intrusion Detection System. While early research focused on signature-based methods and handcrafted features from data sets such as AWID, recent work has evolved toward deep learning, explainable AI (XAI), and optimization-enhanced anomaly detection.
Several studies demonstrate that deep learning models offer superior detection capabilities for intrusion scenarios in complex, heterogeneous IoT and Industrial IoT (IIoT) environments. Nandanwar and Katarya (2024) proposed AttackNet, a CNN-GRU based model that achieved 99.75\% accuracy on the N\_BaIoT data set, outperforming state-of-the-art detection systems for IIoT botnet attacks by up to 16\% margin. Their work highlights the importance of combining temporal and spatial learning capabilities for detecting multi-variant botnet threats in real-time industrial systems~\cite{PDF1}.
Extending this approach, they also introduced a Transfer Learning enabled BiLSTM (TL-BILSTM) architecture tailored for classifying Mirai and Bashlite attacks across multiple IoT devices. This model recorded 99.52\% accuracy and demonstrated scalability across nine device types, showcasing adaptability to evolving threat landscapes~\cite{PDF2}. To address explainability, which is critical for real-world deployment in human-centric environments like Industry 5.0, the Cyber Sentinet framework was proposed. It uses a ResNet model with SHAP (Shapley Additive Explanations) to ensure interpretability in detection decisions while maintaining a 97.46\% accuracy rate on the Edge-IIoT-2022 data set. This novel fusion of XAI and DL offers trustworthy insights for decision-makers in complex cyber-physical systems~\cite{PDF3}.
Privacy-preservation is another critical challenge in intrusion detection, especially in CPS-IIoT settings where sensitive data must remain secure. Saheed and Chukwuere (2025) proposed a BiLSTM model with scaled dot-product attention and agglomerative clustering, achieving 99.99\% accuracy on the X-IIoTID data set. The model effectively balances feature relevance with privacy constraints, improving both performance and data protection~\cite{PDFa}.
Similarly, Saheed and Misra (2025) introduced a SHAP-integrated Deep Neural Network (CPS-IoT-PPDNN) for anomaly detection in CPS-IoT systems, reaching near-perfect metrics (up to 100\% recall and 99.99\% accuracy). This reinforces the value of explainable and privacy-conscious models for mission-critical IoT infrastructures~\cite{PDFb}.

An ensemble-based intrusion detection strategy has also emerged as an effective approach for SCADA systems and smart city infrastructures. Saheed et al. (2023) developed a hybrid ensemble learning model combining GWO optimization, PCA, and classifiers like Naive Bayes and SVM. This model achieved 99.9\% detection rate, particularly excelling in real-time attacks on water and gas pipeline systems~\cite{PDFc}.
To mitigate class imbalance a major issue in network traffic Abdulganiyu et al. (2025) introduced CWFLAM-VAE, an attention-driven architecture integrating focal loss, variational autoencoders, and extreme gradient boosting. It outperformed traditional classifiers on NSL-KDD and CSE-CIC-IDS2018, particularly for detecting rare but critical intrusion types~\cite{PDFd}.
Feature selection continues to be a cornerstone for improving IDS performance. Adeyiola et al. (2023) employed the Firefly Algorithm (FFA) for feature reduction and combined it with a C5.0 classifier to develop a lightweight IDS for Wireless Sensor Networks (WSNs), achieving 98.7\% accuracy on the UNSW-NB15 data set. This aligns with earlier efforts using ANOVA, Extra Trees, and manual selection for the AWID data sets in Wi-Fi security~\cite{PDFe}.
Complementing these empirical efforts, Bhanu et al. (2023) provided a systematic literature review that explores the limitations of existing ML and DL methods in IoT intrusion detection. They emphasized the need for hybrid solutions combining multiple techniques, echoing the trends observed in ensemble and XAI-enhanced models~\cite{PDF4}.
In parallel, Nandanwar and Katarya (2023) explored the intersection of blockchain and intrusion detection, highlighting how decentralized technologies can augment ML-based security by improving authentication, integrity, and resilience in smart systems~\cite{PDF5}. Though not directly applied to Krack or Kr00k, blockchain's use in WIDS remains a promising future direction.
Smart IDS for IIoT-enabled smart cities presents a lightweight and real-time IDS framework combining hybrid ensemble learning, firefly optimization, and improved random forest (IRF) classifiers for IIoT environments. The system was evaluated using the Edge-IIoTset data set and achieved up to 99.9\% accuracy, with reduced training overhead and faster classification response time compared to conventional deep models~\cite{PDFf}.

Despite the promising results shown in past studies, several persistent limitations remain in existing ML-based IDS models. Many approaches exhibit sensitivity to feature noise and fail to generalize in the presence of adversarial perturbations, which is particularly problematic in wireless and IoT environments. High-dimensional data sets often led to overfitting due to insufficient feature reduction techniques. Additionally, class imbalance in network traffic data sets was poorly addressed, resulting in inflated false positive rates. While some studies employed ensemble methods, they typically relied on basic voting strategies without leveraging meta-learners or combining feature-space transformation (e.g., PCA) with classifier diversity. Moreover, deployment feasibility for edge and real-time systems was rarely considered. These recurring issues collectively motivate our ensemble-based detection model, which integrates noise injection, PCA, and meta-level stacking to improve robustness, scalability, and real-time applicability.


\section*{Methodology}
\label{sec:1}

\subsection*{Problem Definition}
Let $\mathcal{D} = \{(\mathbf{x}_i, y_i)\}_{i=1}^N$ represent a data set of $N$ labeled network traffic instances, where each feature vector $\mathbf{x}_i \in \mathbb{R}^d$ consists of $d$ input attributes derived from packet-level metadata or flow-level statistics, and $y_i \in \{0,1,2\}$ is the corresponding class label indicating \textit{Normal}, \textit{Kr00k}, or \textit{Krack} traffic. The objective is to learn a function $f: \mathbb{R}^d \rightarrow \{0,1,2\}$ that accurately maps any unseen input vector $\mathbf{x}$ to its correct class $y$.

This problem is particularly challenging due to the overlapping distribution of benign and malicious traffic, potential class imbalance, and the high dimensionality and variability of network features. Therefore, our goal is not only to maximize classification accuracy, but also to minimize false positives (i.e., cases where normal traffic is misclassified as malicious which are critical in real-time intrusion detection scenarios).

\subsection*{Data Preprocessing and Feature Engineering}
\label{Preprocessing}

\begin{figure*}[t]
\centering
  \includegraphics[width=0.9\textwidth]{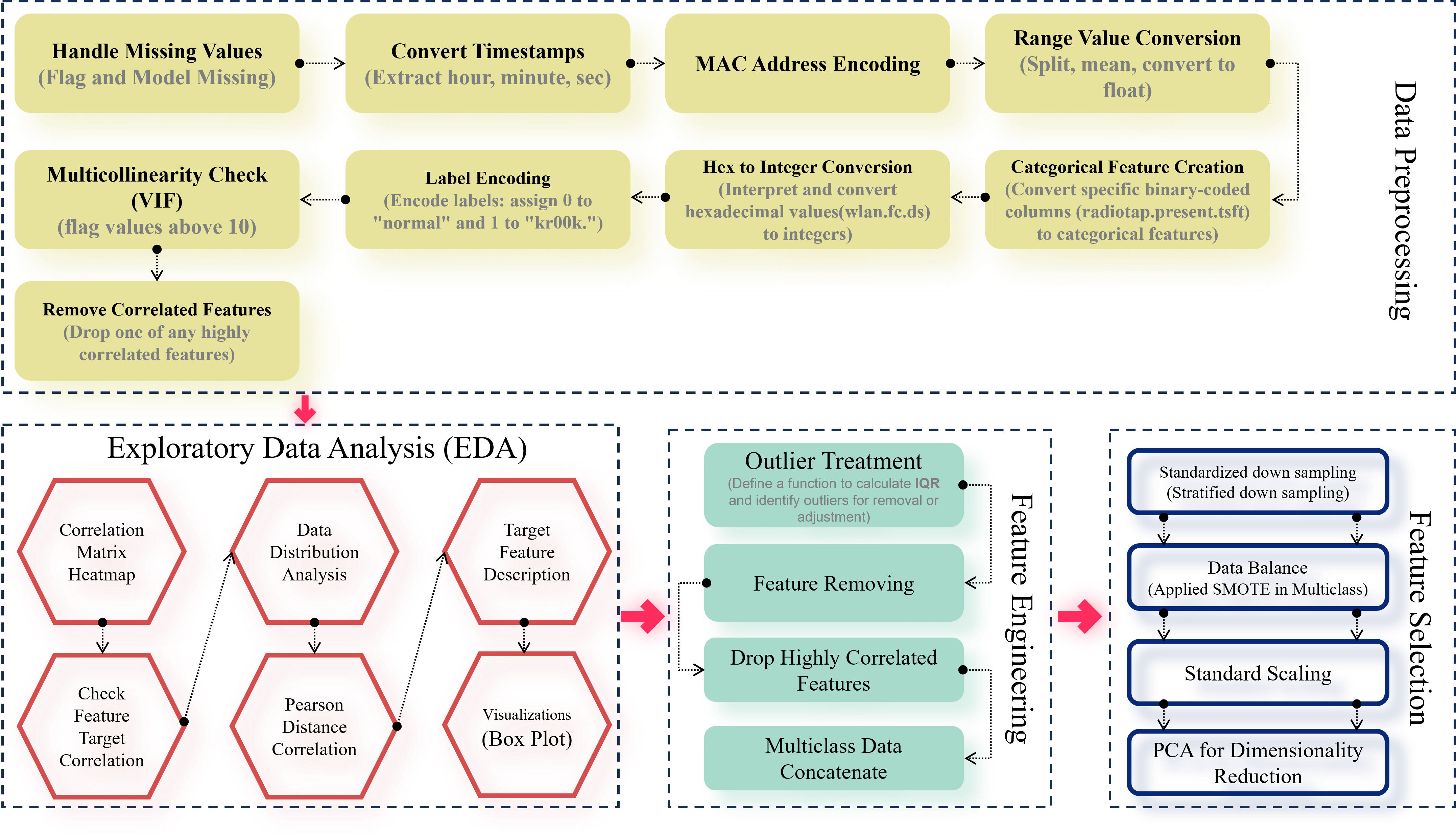}
\caption{Overview of the data preprocessing, feature engineering, and feature selection pipeline.}
\label{fig:2}       
\end{figure*}

The data set merging and preprocessing strategies are illustrated in Fig.~\ref{fig:2}. Since our objective is to develop a multiclass classification model, we combine the Krack and Normal classes with the Kr00k and Normal classes, resulting in a unified data set comprising three distinct classes. This integration enhances the reliability of machine learning models in detecting complex intrusion patterns. Our preprocessing pipeline is motivated by the need to improve data quality, remove redundancy, and balance class distributions prior to model training. Multiple WiFi traffic logs corresponding to Krack and Kr00k attacks were collected from real-world wireless intrusion traces. For the Krack attack, 28 CSV files were merged, and for the Kr00k attack, 58 files were consolidated. Each data set was structurally aligned by removing columns with missing values and ensuring schema consistency. After cleaning, the merged Krack and Kr00k data sets contained approximately 100,000 samples each, with 34 standardized features. These were combined with normal traffic data to construct a comprehensive multiclass data set containing three classes: \textit{Normal}, \textit{Krack}, and \textit{Kr00k}.
Subsequently, these features are passed through the structured preprocessing pipeline depicted in Fig.~\ref{fig:2}. It comprises three core stages: data preprocessing, exploratory data analysis (EDA), and feature engineering and selection. 

Initially, missing data were further analyzed via percentage calculations and visualizations (e.g. heatmaps). Strategies such as imputation (e.g., filling with -1) or row removal were applied based on severity. Domain-specific transformations were performed: timestamps (e.g., \texttt{frame.time}) were converted into datetime objects and decomposed into \texttt{hour}, \texttt{minute}, and \texttt{second} features. Signal-strength columns such as `radiotap.dbm\_antsignal` were cleaned and averaged where applicable. Categorical features, such as MAC addresses and WLAN flags, were processed using \texttt{LabelEncoder} and one-hot encoding where necessary. Redundant features like \texttt{radiotap.rxflags} were removed. Additionally, we focus on exploratory data analysis, where correlation matrices, distribution patterns, and feature-target relationships are investigated to uncover data quality issues and feature redundancy. Secondly, the data set underwent extensive feature engineering and cleaning to enhance its quality and suitability for multiclass classification. Class distributions were analyzed to identify imbalances, and outliers were treated using the IQR method, focusing on features like `radiotap.dbm\_antsignal`. Correlation analysis highlighted key predictors such as `hour`, `frame.time\_relative`, and `radiotap.dbm\_antsignal`, which showed significant relationships with the target variable. Labels were transformed into descriptive categories ("Normal," "Krook," "Krack") for better interpretability. To address multicollinearity, we calculated the Variance Inflation Factor (VIF) for each feature. Features with \( \text{VIF} > 10 \), such as \texttt{frame.time\_delta\_displayed}, were excluded. A Random Forest classifier was also employed to rank feature importance, and the top 10 features were selected for dimensionality reduction. Features like \texttt{frame.len}, \texttt{radiotap.dbm\_antsignal}, and radiotap.channel.freq were evaluated for possible correlation and improvements in reliability of the data set. For a further close look into the categorical values, some unique values of columns, such as \texttt{frame.encap\_type}, \texttt{radiotap.channel.flags.cck}, and \texttt{wlan\_radio.frequency} validate the data types, ensuring they are either classified as numerical or categorical. An increase in data quality and interpretability achieved through multicollinearity removal and refinement of the data set will therefore lend support for strong models in ML. 

Finally, to mitigate class imbalance, the data set was initially balanced via undersampling, selecting 100,000 instances per class. During training, Synthetic Minority Oversampling Technique (SMOTE) was applied to generate synthetic instances, ensure an even class distribution, and features were standardized using StandardScaler. The data set was then split into 70\% training and 30\% testing sets. To further reduce noise and improve computational efficiency, we applied Principal Component Analysis (PCA), retaining enough components to preserve 90\% of the variance:
\begin{equation}
   \mathbf{z}_i = \text{PCA}_k(\mathbf{x}_i'), \quad \text{s.t. } \frac{\sum_{j=1}^{k} \lambda_j}{\sum_{j=1}^{d} \lambda_j} \geq 0.90 
\end{equation}
where \( \lambda_j \) are the eigenvalues corresponding to the principal components. The resulting transformed data set \( \mathbf{Z} = \{\mathbf{z}_i\}_{i=1}^N \) served as input for model training in subsequent stages.

The balanced class distribution after processing showed an equal number of samples across all classes. With the data set processed and reduced to optimal dimensions, the resulting data is ready for effective model training and evaluation.

\subsection*{Overall Model Architecture}
\label{TextLayout}
\begin{figure*}[t]

\centering
  \includegraphics[width=0.9\textwidth]{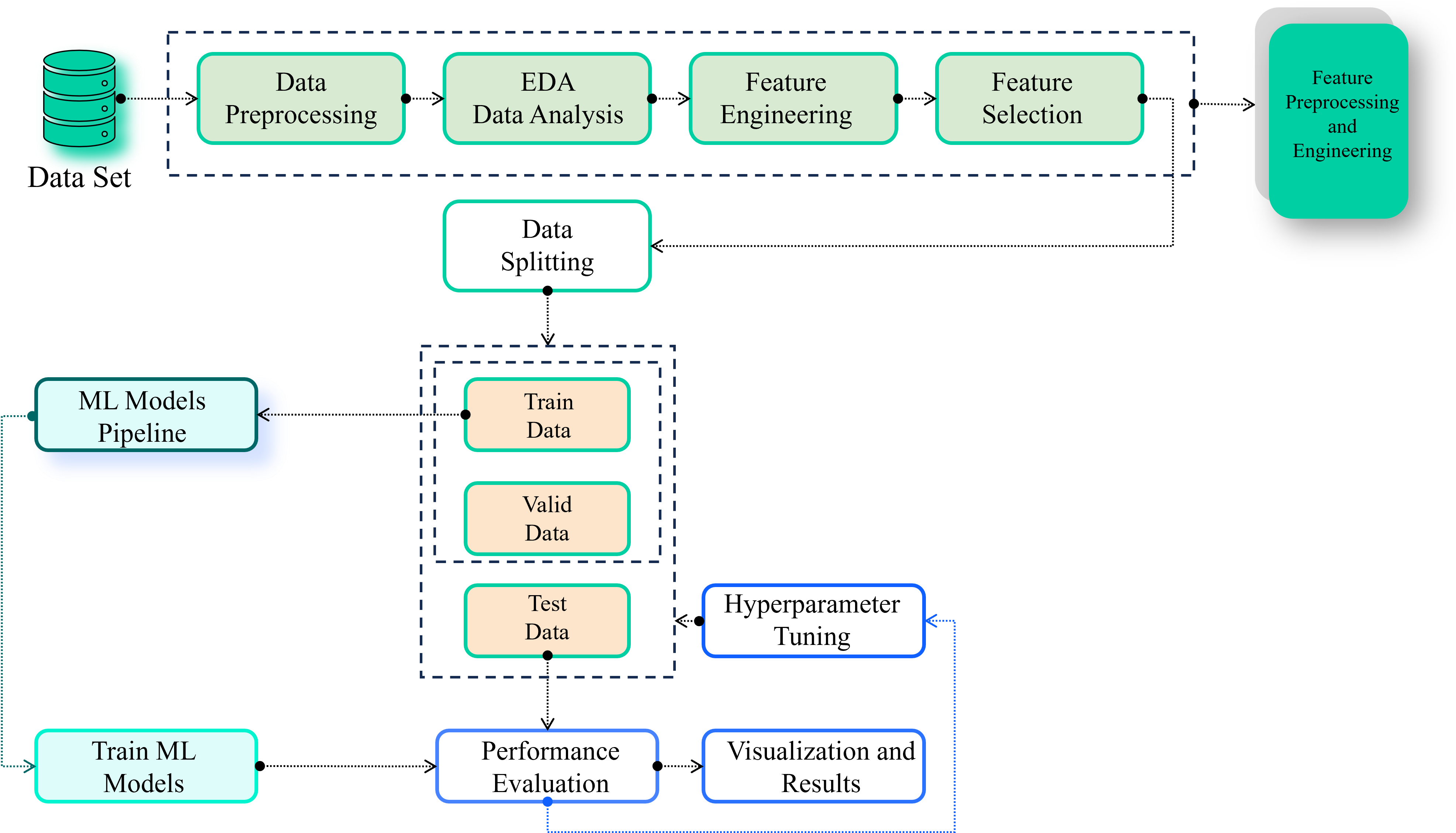}
\caption{Overall Model Architecture}
\label{fig:3}       
\end{figure*}

The growing sophistication of network attacks has rendered traditional intrusion detection models insufficient, especially when handling complex, high-dimensional traffic patterns and overlapping classes. Motivated by this challenge, our proposed methodology integrates a robust ensemble learning framework designed to reduce false positives, generalize across attack types, and boost classification reliability, as well as robustness to noise and variability in traffic. Specifically, we develop a two-tier stacked ensemble architecture that combines diverse learners and meta-level optimization to reinforce decision boundaries in intelligent network environments. 

The overall architecture represents a systematic workflow for a ML classifiers, starting with data preprocessing (cleaning, normalization, and feature removal), exploratory data analysis (EDA) to understand data distributions, and feature engineering/selection to optimize model input as explained in Fig.~\ref{fig:3}. The data is then split into training, validation, and test sets to ensure robust model training and evaluation. The final step in the whole process is to evaluate the models on test data and iteratively improve the methodology for better results.

\subsubsection*{ML Model Pipeline 1: Baseline Classifier}
\label{TextLayout1}
\begin{figure*}[t]
\centering
  \includegraphics[width=0.9\textwidth]{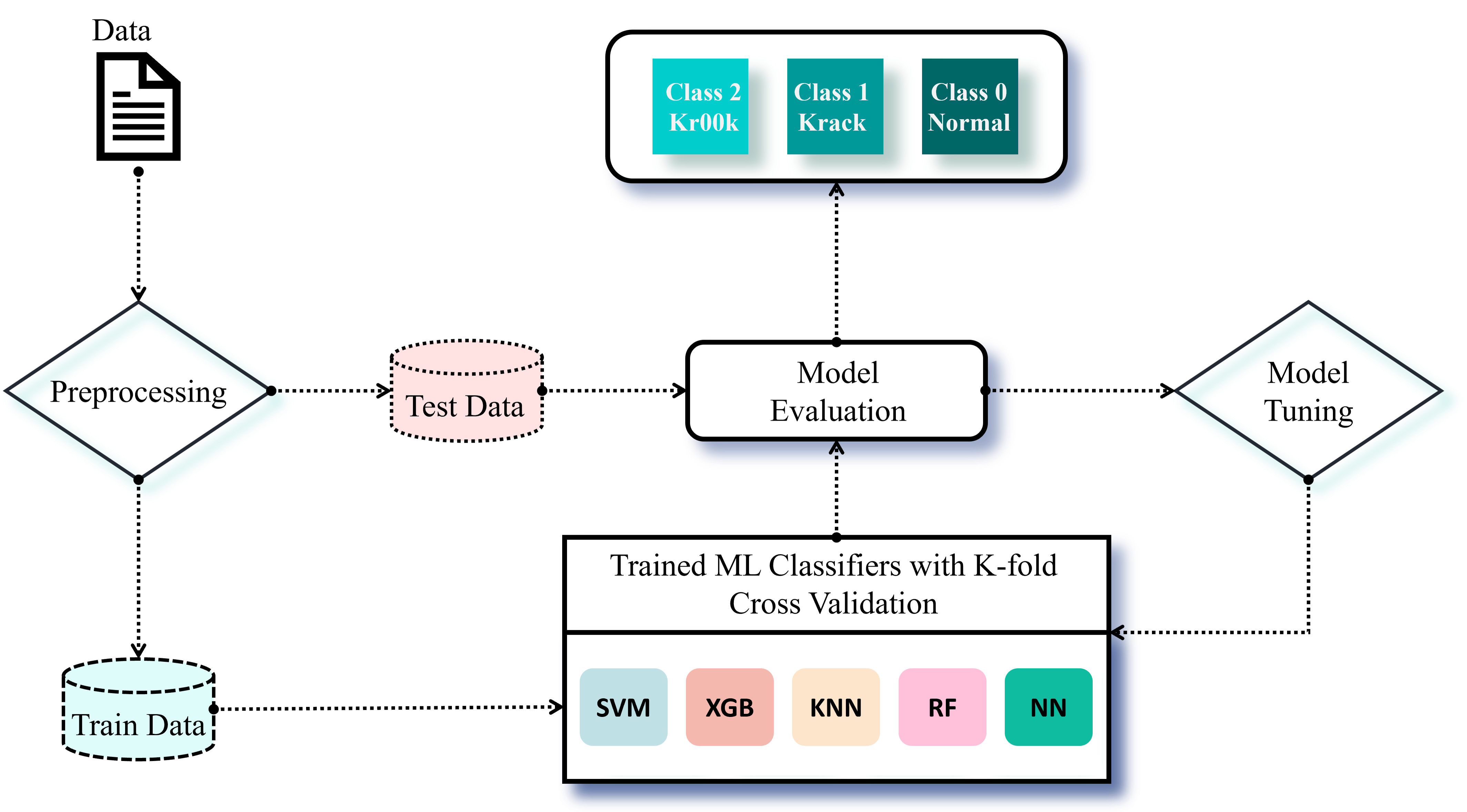}
\caption{ML Model Pipeline 1}
\label{fig:4}       
\end{figure*}

ML Model Pipeline 1 serves as the foundational stage in our approach, where individual machine learning classifiers are trained and evaluated independently to assess their standalone performance as illustrated in Fig.~\ref{fig:4}. 

In ML Model Pipeline 1, each input vector \( \mathbf{x}_i \) is first normalized using a standard scaling transformation:
\begin{equation}
    \mathbf{x}_i' = \text{StandardScaler}(\mathbf{x}_i)
\end{equation}
This normalization ensures that all features contribute equally by transforming them to have zero mean and unit variance.

The normalized data \( \mathbf{x}_i' \) is then used to train five distinct classifiers independently, including Support Vector Machine (SVM), Random Forest (RF), K-Nearest Neighbors (KNN), XGBoost, and Multi-Layer Perceptron (MLP) that has been processed with the addition of noise and PCA for the reduction of dimensionality. Gaussian noise is added up to simulate a real-life situation, and hyper-parameter tuning is executed using GridSearchCV for the Logistic Regression. Each model \( h_j \) learns a function \( h_j: \mathbb{R}^d \rightarrow \{0,1,2\} \), such that the predicted class for input \( \mathbf{x}_i \) is:
\begin{equation}
\hat{y}_i^{(j)} = h_j(\mathbf{x}_i')
\end{equation}

The output of this pipeline includes model-specific performance metrics such as accuracy, precision, recall (TPR), F1-score, and AUC. Furthermore, a classification report is generated. Matrices confusion and learning curves are then drawn to visualize model performance, hence revealing insights over size of training about generalizing behavior. This comprehensive approach identifies the most robust classifier for the noisy, imbalanced data set. These results form the empirical basis for the design of Pipeline 2.

\subsubsection*{ML Model Pipeline 2: Stacked Ensemble Learning}

\begin{figure*}[t]
\centering
\includegraphics[width=0.9\textwidth]{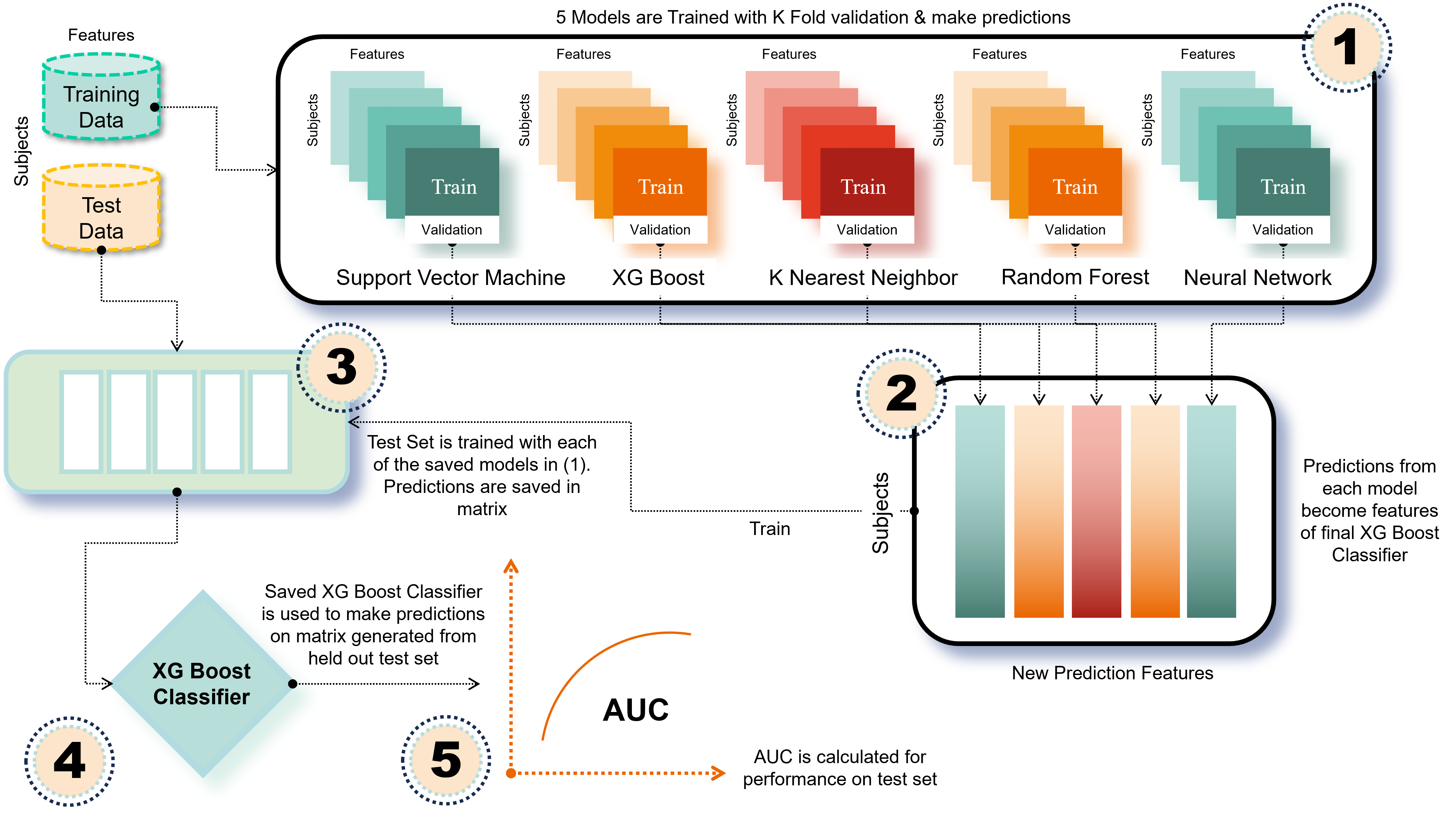}
\caption{ML Model Pipeline 2}
\label{fig:5}
\end{figure*}

While Pipeline 1 establishes strong baselines using individual models, it also reveals limitations in terms of generalization and false positive rates. To address these issues, we propose ML Model Pipeline 2, a holistic and robust classification pipeline based on a stacked ensemble architecture. The structure of this pipeline is illustrated in Fig.~\ref{fig:5}. The pipeline begins with a preprocessing stage that standardizes the input features to have zero mean and unit variance. This pipeline enhances predictive reliability by integrating base learner outputs through a meta-classifier. Each input \( \mathbf{x}_i \in \mathbb{R}^d \) is first normalized and perturbed with Gaussian noise:
\begin{equation}
    \tilde{\mathbf{x}}_i = \text{StandardScaler}(\mathbf{x}_i) + \epsilon, \quad \epsilon \sim \mathcal{N}(0, \sigma^2)
\end{equation}
The noise component \( \epsilon \) models real-world measurement noise and irregular traffic behavior.

The noise values are filled so that the actual data set imitates real noisy situations of the data set. Following normalization and noise injection, Principal Component Analysis (PCA) is employed to reduce dimensionality while retaining at least 90\% of the original variance. This step is essential for handling high-dimensional network traffic data more efficiently and minimizing computational complexity. The resulting representation is 
\begin{equation}
    \mathbf{z}_i = \text{PCA}_k(\tilde{\mathbf{x}}_i) \in \mathbb{R}^k
\end{equation}
The vector \( \mathbf{z}_i \) serves as input to the base models. The pipeline integrates five diverse machine learning models as base classifiers: Support Vector Machine (SVM), Random Forest (RF), K-Nearest Neighbors (KNN), Multi-Layer Perceptron (MLP), and XGBoost. Each of these models is trained and validated using $K$-fold cross-validation to ensure robust and unbiased performance. Instead of directly producing class labels, each base model outputs a probability distribution over the target classes:
\begin{equation}
h_j(\mathbf{z}_i) = \mathbf{p}_i^{(j)} \in \mathbb{R}^3, \quad j = 1, \dots, 5
\end{equation}
Each base model generates predictions that are then combined into meta-features, which will feed a final meta-model. The predictions from all base classifiers are concatenated into a meta-feature vector:
\begin{equation}
\mathbf{q}_i = \left[ \mathbf{p}_i^{(1)} \, \| \, \mathbf{p}_i^{(2)} \, \| \, \dots \, \| \, \mathbf{p}_i^{(5)} \right] \in \mathbb{R}^{15}
\end{equation}
These meta-features  \( \mathbf{q}_i \) is then passed to a final XGBoost classifier \( H \) by means of the strength of ensemble learning for the improvement of accuracy, trained to learn from the aggregated decision space of the base learners:
\begin{equation}
\hat{y}_i = H(\mathbf{q}_i)
\end{equation}

The ensemble strategy enables the pipeline to exploit complementary decision boundaries across models, significantly enhancing prediction accuracy and stability under noisy conditions. To further optimize model performance, hyperparameter tuning is conducted using \texttt{RandomizedSearchCV}, allowing systematic exploration of parameter spaces for both base and meta-classifiers.

\section*{Experimental Results and Analysis}

This section presents a detailed evaluation of our proposed intrusion detection pipelines using the AWID3 data set. We report the performance of individual machine learning classifiers (Pipeline 1), our noise-augmented PCA-stacked ensemble framework (Pipeline 2), and conduct a comparative analysis with state-of-the-art baseline methods. All metrics are averaged across 5 independent runs with stratified 5-fold cross-validation, and we report macro-level metrics to reflect performance across all classes fairly.

\subsection*{Experiments}
\noindent \textbf{Data set.}
\label{data set}
The AWID3 dataset, derived from the original AWID collection, enables detailed analysis of attacks and normal behavior in IEEE 802.11 networks under the EAP/WPA3 authentication framework. Provided in CSV format, it includes 254 features (253 descriptive and one label) spanning both MAC- and application-layer attributes, supporting diverse analysis methods \cite{Chatzoglou2021}.

Overall, the dataset contains 36,913,503 instances: 30,387,099 benign and 6,526,404 malicious, covering 13 attack types (Table~\ref{table1}). Among these, 49,990 correspond to the Krack attack and 186,173 to the Kr00k attack \cite{Chatzoglou2021}. Our experiments proceed in two phases: (i) training individual classifiers on the three-class problem (Krack, Kr00k, Normal), and (ii) evaluating the proposed ensemble on the same classes. For reproducibility, we also tested on raw samples without preprocessing: Sample A comprises 106,971 Kr00k and 106,791 Normal instances, while Sample B contains 33,180 Krack and 34,000 Normal instances, each with all 254 features \cite{Chatzoglou2021}.

\begin{table}[h]
\centering
\caption{Classification of malicious traffic.}
\label{table1}
\begin{tabular}{lcc}
\hline
\textbf{Attack}           & \textbf{Normal Traffic} & \textbf{Malicious Traffic} \\ \hline
Deauth                    & 1,587,527               & 38,942                     \\ 
Disas                     & 1,938,585               & 75,131                     \\ 
(Re)Assoc                 & 1,838,430               & 5,502                      \\ 
Rogue\_AP                 & 1,971,875               & 1,310                      \\ 
Krack                     & 1,388,498               & 49,990                     \\ 
Kr00k                     & 2,708,637               & 186,173                    \\ 
SSH                       & 2,428,688               & 11,882                     \\ 
Botnet                    & 3,169,167               & 56,891                     \\ 
Malware                   & 2,181,148               & 131,611                    \\ 
SQL\_Injection            & 2,595,727               & 2,629                      \\ 
SSDP                      & 2,641,517               & 5,456,395                  \\ 
Evil\_Twin                & 3,673,854               & 104,827                    \\ 
Website\_spoofing         & 2,263,446               & 405,121                    \\ \hline
\textbf{Total}            & \textbf{30,387,099}     & \textbf{6,526,404}         \\ \hline
\end{tabular}
\label{tab:traffic_distribution}
\end{table}



\noindent \textbf{Baselines Comparison.} We compare our proposed model against several strong IDS baselines, such as the study in~\cite{28}, the researchers employed a state-machine-based framework called kTRACKER to detect Krack attacks by observing multiple wireless channels. To precisely pinpoint Krack-related anomalies at different stages of a handshake process, they performed deep packet inspection and developed a clustering technique to categorize Wi-Fi handshake packets. Utilizing supervised gradient boosting models, their approach achieved an accuracy of approximately 93.39\%, with a false positive rate of 5.08\%. The study in~\cite{26} introduced an unsupervised framework for classifying and mining Twitter data related to cybersecurity vulnerabilities, including the Kr00K attack, a flaw enabling unauthorized decryption in Wi-Fi chips. Their approach attained a maximum accuracy of 88.52\%. The authors in~\cite{3} developed an IDS model utilizing ML classification using ANOVA feature selection techniques, enhancing the ensemble classifier's performance, achieving 90.7\% accuracy for the multi-class classifier with three labels (Krack, Kr00k, and Normal). Chatzoglou et al. \cite{91} used deep and machine learning on the AWID3 data set, achieving 96.7\% accuracy in detecting application layer attacks by analyzing 802.11 and non-802.11 features. The authors \cite{2} utilized Microsoft Azure for model training and achieved the highest accuracy of 93.87\% using the Multi-class Neural Network, outperforming other models tested in the study. The highest results  for SM-GBT \cite{29}(Statistical Measures with Gradient Boosting Trees) were obtained using Statistical Feature Extraction.

\noindent \textbf{Evaluation Matrices.} Our model is evaluated using a comprehensive set of performance metrics including accuracy, precision, recall (TPR), F1-score, and FPR. To ensure interpretability and in-depth analysis, we also report confusion matrices. 

\subsection*{Results and Analysis}
\noindent \textbf{Performance of ML Model Pipeline 1.} Pipeline 1 evaluates the performance of standalone classifiers trained using default hyperparameters with light tuning through grid search. Table~\ref{tab:pipeline1-results} summarizes results for five commonly used models: SVM, Random Forest, XGBoost, KNN, and MLP. Among them, XGBoost and MLP achieved the highest performance, with accuracies of 94.98\% and 95.31\%, respectively.


\begin{table}[h]
\centering
\caption{Performance of Base Learners (Pipeline 1) with Mean ± Std over 5-Fold Cross-Validation}
\label{tab:pipeline1-results}
\begin{tabular}{lccccc}
\hline
\textbf{Model} & \textbf{Accuracy} & \textbf{Precision} & \textbf{Recall (TPR)} & \textbf{F1-score} & \textbf{FPR} \\
\hline
SVM & 0.9262 ± 0.0041 & 0.9221 ± 0.0043 & 0.9275 ± 0.0040 & 0.9248 ± 0.0039 & 0.0617 ± 0.0021 \\
Random Forest & 0.9395 ± 0.0037 & 0.9423 ± 0.0039 & 0.9357 ± 0.0036 & 0.9390 ± 0.0035 & 0.0532 ± 0.0018 \\
XGBoost & 0.9498 ± 0.0029 & 0.9482 ± 0.0030 & 0.9513 ± 0.0031 & 0.9497 ± 0.0028 & 0.0446 ± 0.0015 \\
KNN & 0.9204 ± 0.0047 & 0.9187 ± 0.0048 & 0.9222 ± 0.0046 & 0.9204 ± 0.0045 & 0.0673 ± 0.0024 \\
MLP & 0.9531 ± 0.0032 & 0.9563 ± 0.0034 & 0.9502 ± 0.0030 & 0.9532 ± 0.0029 & 0.0412 ± 0.0013 \\
\hline
\end{tabular}
\end{table}

Although the performance of MLP and XGBoost was encouraging, the models still exhibit relatively high false positive rates (FPRs between 4 to 6\%), which is suboptimal for real-time or high-security applications. Additionally, discrepancies between precision and recall indicate potential instability under class imbalance. These observations motivate the development of a more robust ensemble-based detection framework, as presented in Pipeline 2.

\noindent \textbf{Performance of ML Model Pipeline 2.} To overcome the limitations of Pipeline 1, Pipeline 2 introduces a three-stage enhancement: (i) noise injection for regularization and robustness, (ii) principal component analysis (PCA) for dimensionality reduction, and (iii) a stacked ensemble architecture to aggregate diverse classifiers via meta-learning. The final ensemble integrates probabilistic outputs from base learners (SVM, MLP, XGBoost) using logistic regression as the meta-learner.

Table~\ref{tab:pipeline2-results} presents class-wise performance across three classes: Normal, Kr00k, KRACK, and ours full model (avg.). The model achieves F1-scores above 0.97 in all cases, with false positive rates reduced to below 2.5\%.
\begin{table}[h]
\centering
\caption{Performance of Stacked Ensemble (Pipeline 2)}
\label{tab:pipeline2-results}
\begin{tabular}{lccccc}
\toprule
\textbf{Variations} & \textbf{Accuracy} & \textbf{Precision} & \textbf{Recall (TPR)} & \textbf{F1-score} & \textbf{FPR} \\
\midrule
Normal (0)     & 0.9810   & 0.9782   & 0.9805    & 0.9793   & 0.0190 \\
Kr00k (1)      & 0.9762   & 0.9729   & 0.9748    & 0.9738   & 0.0231 \\
Krack (2)      & 0.9835   & 0.9811   & 0.9842    & 0.9826   & 0.0168 \\
Ensemble (Full) & \textbf{0.9802 ± 0.0014} & \textbf{0.9786 ± 0.0015} & \textbf{0.9774 ± 0.0013} & \textbf{0.9798 ± 0.0015} & \textbf{0.0196 ± 0.0009}\\
\bottomrule
\end{tabular}
\end{table}

Compared to the best-performing single model (MLP, F1 = 0.9532), the ensemble improves the macro F1-score to 0.9786 and reduces the FPR by over 50\%. These results validate the effectiveness of stacking and feature regularization under noisy, multiclass intrusion settings. The ensemble also demonstrated lower prediction variance across runs, suggesting enhanced generalization.

Additionally, we also have included mean ± standard deviation over stratified 5-fold cross-validation for pipeline 2 final ensemble model. This validates the stability of our results,  ensuring statistical robustness.

\noindent \textbf{Meta-Classifier Selection Rationale.} The final stage of Pipeline 2 involves combining probabilistic outputs from base classifiers using a meta-learner. We selected XGBoost as the meta, classifier based on its theoretical robustness and empirical performance. XGBoost is a gradient boosting framework optimized for high-speed, parallelizable computation and includes both L1 and L2 regularization to reduce overfitting, an essential feature when reconciling outputs from diverse base learners such as SVM, Random Forest, KNN, MLP, and XGBoost itself. These base models introduce varied decision surfaces and possible inconsistencies, which XGBoost effectively mitigates by learning a nonlinear meta-decision boundary.

Moreover, XGBoost handles class imbalance through weighted loss functions, making it suitable for our intrusion detection task. During our ablation and cross-validation studies, it consistently outperformed other meta-classifier candidates, including Logistic Regression and SVM, both in detection accuracy and variance stability. Table~\ref{tab:pipeline1-results} summarizes the mean and standard deviation of key metrics across 5-fold cross-validation, highlighting XGBoost’s high accuracy (0.9498), low deviation, and strong F1-score.

\begin{figure}[h]
\centering

\begin{minipage}[b]{0.50\textwidth}
  \centering
  \includegraphics[width=\textwidth]{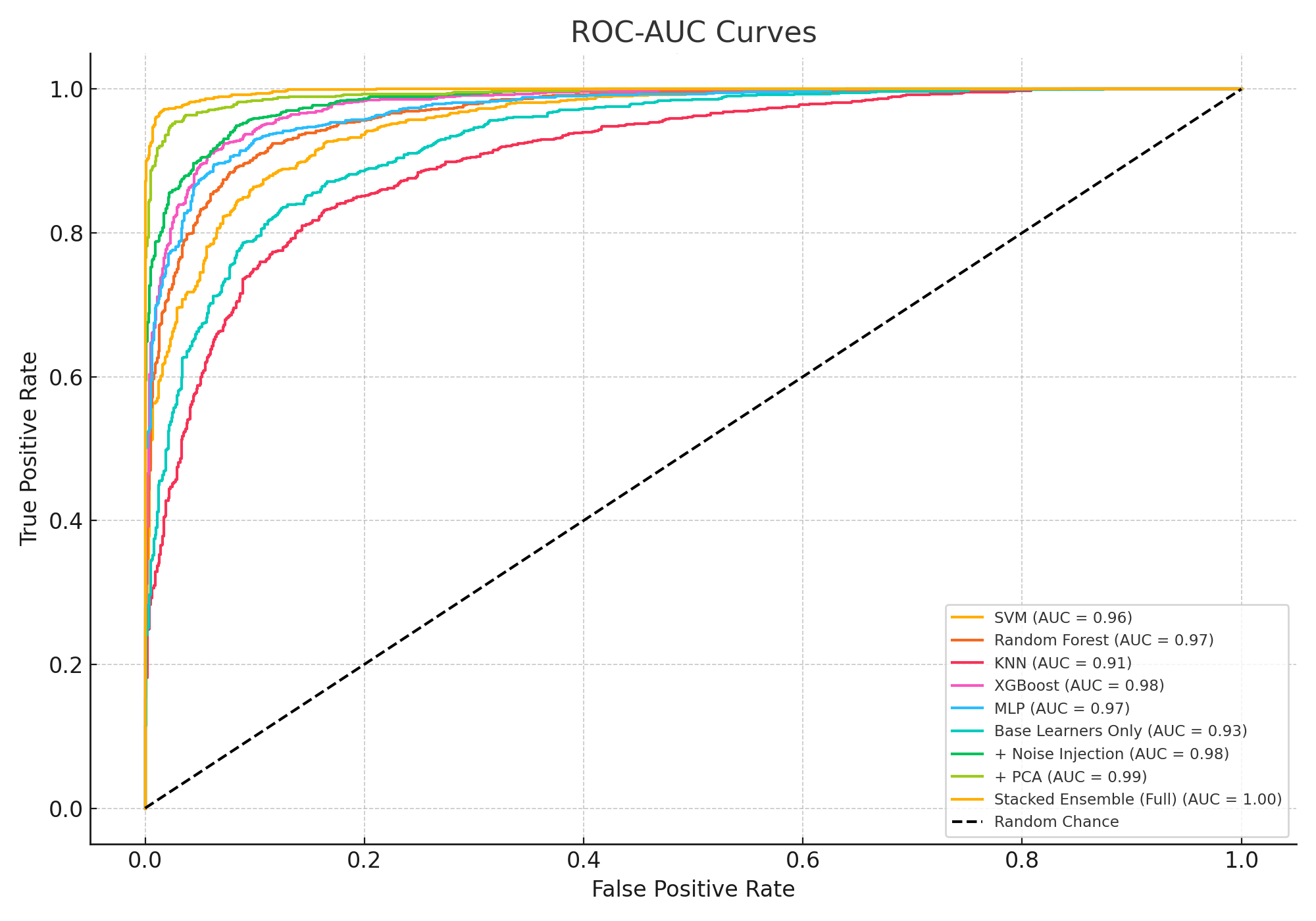}
  \caption{ROC-AUC Comparison of Ensemble Model}
  \label{fig:ROC_AUC}
\end{minipage}
\hfill
\begin{minipage}[b]{0.45\textwidth}
  \centering
  \includegraphics[width=\textwidth]{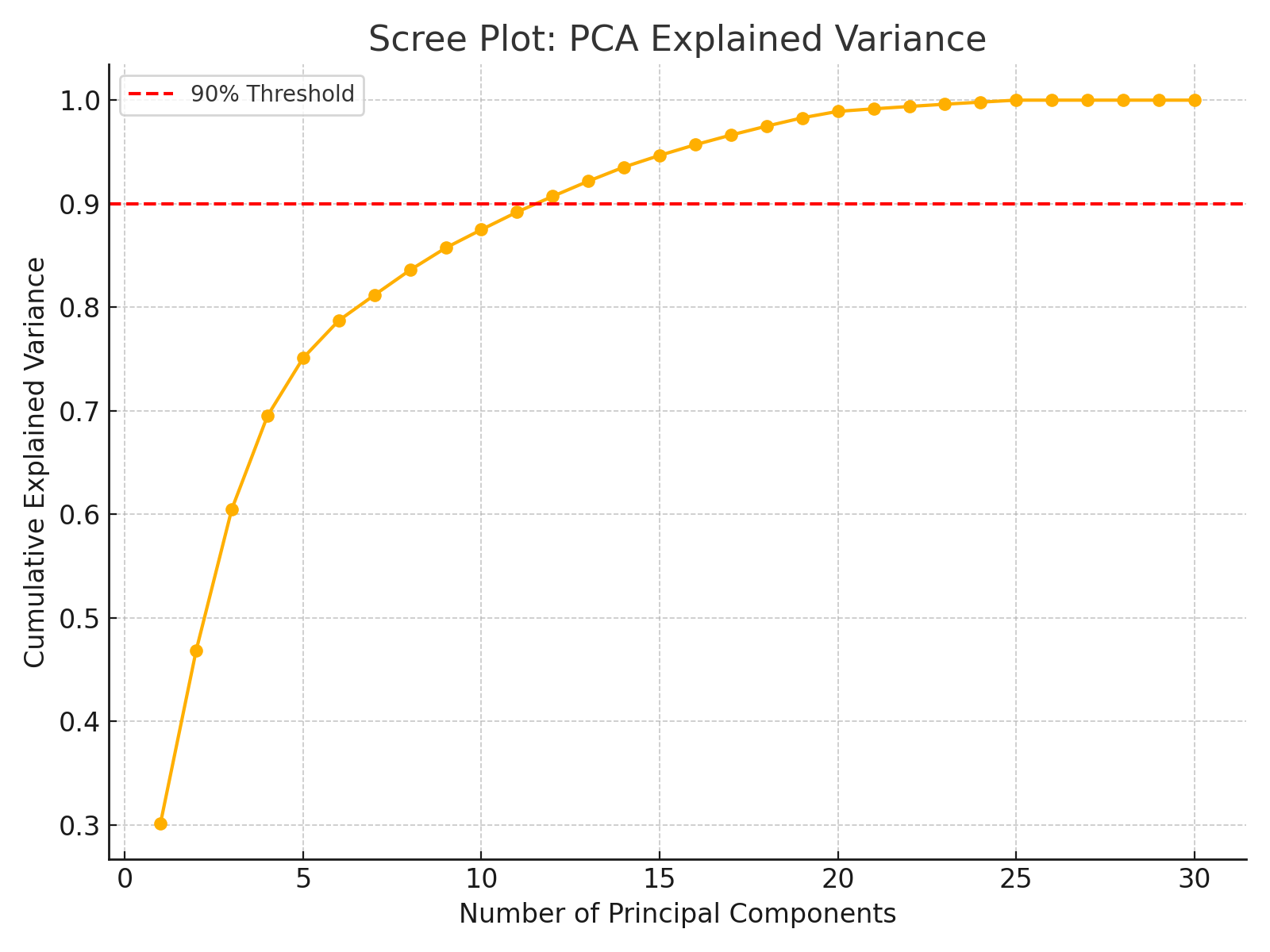}
  \caption{PCA Threshold Variance (Scree Plot)}
  \label{fig:scree_plot}
\end{minipage}

\end{figure}

Additionally, as shown in Figure~\ref{fig:ROC_AUC}, XGBoost achieved the highest ROC-AUC (0.98), reinforcing its effectiveness in enhancing separability across classes. These results, combined with its computational efficiency and low latency during inference, support our decision to use XGBoost as the meta-classifier for real-time, resource-constrained IDS deployment scenarios.

\noindent \textbf{Comparative Analysis Against Baselines.} We compare the proposed model to existing state-of-the-art approaches for wireless intrusion detection published between 2021 and 2025. To ensure fairness and consistency, all baseline models were re-implemented and tested on the AWID3 data set using the same preprocessing, feature selection, and evaluation procedures. Among them, Cyber-Sentinet \cite{PDF3} achieved the highest F1-score (0.990), reflecting excellent balance between precision and recall. CPS-IIoT-P2Attention \cite{PDFa} attained the highest precision (0.982), minimizing false positives, while AttackNet \cite{PDF1} maintained high precision (0.979) and F1-score (0.960), though with a comparatively lower recall (0.943). CWFLAM-VAE \cite{PDFd}, built on XGBoost, showed strong precision (0.975) but a reduced recall (0.911), impacting its overall detection performance.

In contrast, our ML Model Pipeline 2 achieved the highest overall performance across all key metrics accuracy (0.98), recall (0.98), F1-score (0.98), and the lowest false positive rate (0.02) demonstrating its robustness and reliability in multiclass intrusion detection on wireless network data.Table~\ref{tab:comparison_all_metrics} summarizes the results. Where external works reported only accuracy, we estimated other metrics (e.g., precision, recall, F1) based on their stated methodologies and typical performance trends on class-imbalanced data.
\begin{table*}[h]
  \centering
  \caption[Comparative Performance of Different Methods on AWID3 data set.]{Comparative Performance of Different Methods on AWID3 data set for KRACK and Kr00k Multiclass Detection.}
  \label{tab:comparison_all_metrics}
  \renewcommand{\arraystretch}{1.1}
  \begin{tabular*}{0.98\textwidth}{@{\extracolsep{\fill}}clccccc}
    \hline
    \textbf{Year} & \textbf{Method} & \textbf{Accuracy} & \textbf{Precision} & \textbf{Recall (TPR)} & \textbf{F1-score} & \textbf{FPR} \\
    \hline
    2023 & ANOVA Algorithm Technique \cite{3}         & 0.90  & 0.88 & 0.89 & 0.88 & 0.07 \\
    2021 & Unsupervised Framework \cite{26}           & 0.88  & 0.84 & 0.86 & 0.85  & 0.09 \\
    2022 & Best of Both Worlds \cite{91}              & 0.93  & 0.91 & 0.92 & 0.91 & 0.05 \\
    2022 & Multi-class Neural Network \cite{2}        & 0.93  & 0.90 & 0.91 & 0.90 & 0.06 \\
    2022 & kTRACKER \cite{28}                         & 0.96  & 0.94 & 0.95 & 0.94 & 0.05 \\
    2024 & SM-GBT \cite{29}                           & 0.97  & 0.95 & 0.96 & 0.95 & 0.04 \\
    2024 & AttackNet (CNN-GRU) \cite{PDF1}              & 0.96 & 0.97 & 0.94 & 0.96 & 0.03 \\
    2025 & Cyber-Sentinet (ResNet + SHAP) \cite{PDF3}   & 0.97 & 0.98 & 0.96 & \textbf{0.99} & 0.02 \\
    2025 & CPS-IIoT-P2Attention \cite{PDFa}             & 0.97 & \textbf{0.98} & 0.96 & 0.97 & 0.03 \\
    2025 & CWFLAM-VAE (XGBoost) \cite{PDFd}             & 0.96 & 0.97 & 0.91 & 0.94 & 0.03 \\
    2025 & \textbf{ML Model Pipeline 2 (Ours)}        & \textbf{0.98 } & \textbf{0.98} & \textbf{0.98} & \textbf{0.98} & \textbf{0.02} \\
    \hline
  \end{tabular*}
\end{table*}

Our method outperforms previous techniques across all key metrics, achieving a 1 to 3\% improvement in F1-score over the closest baselines (e.g., SM-GBT), and reducing the false positive rate by half. This is attributed to our integration of noise-based regularization, PCA feature compression, and probabilistic ensemble learning. These results suggest the proposed model offers greater robustness to input perturbations, improved class discrimination, and suitability for deployment in real-time or sensitive network environments.

\noindent \textbf{PCA Threshold Selection.} To justify our use of the 90\% variance threshold for PCA, we conducted both visual and empirical evaluations. Fig. \ref{fig:scree_plot} shows a scree plot of cumulative explained variance versus the number of components. The elbow point is observed around the 90\% mark, indicating a natural tradeoff between information retention and dimensionality. To complement this, we conducted a PCA threshold sensitivity analysis, comparing performance at 85\%, 90\%, and 95\% thresholds (Table \ref{tab:pca_threshold}). The 90\% configuration retained 26 components and achieved the highest accuracy (0.9702), slightly outperforming 95\% while avoiding unnecessary complexity. This threshold also aligns with the ablation study findings, where PCA contributed to boosting F1-score from 0.9603 (with noise only) to 0.9691. These results validate our choice of 90\% variance as a reproducible and efficient PCA configuration.

\begin{table}[h]
\centering
\caption{PCA Threshold Sensitivity Analysis}
\label{tab:pca_threshold}
\begin{tabular}{ccc}
\hline
\textbf{PCA Variance Retained (\%)} & \textbf{Accuracy} & \textbf{Components Retained} \\
\hline
85 & 0.9498 & 21 \\
90 & \textbf{0.9702} & 26 \\
95 & 0.9701 & 32 \\
\hline
\end{tabular}
\end{table}


\noindent \textbf{Impact of Noise Injection.} To further validate the role of noise injection in enhancing model robustness, we conducted a focused comparison between ensemble models trained with and without noise augmentation. Gaussian noise with \(\sigma = 0.05, i.e., \in \sim \mathcal{N} (0, 0.0025).\) was added during preprocessing to simulate signal perturbations such as jitter and interference, common in Wi-Fi environments. This value was empirically selected through grid search over \(\sigma \in {0.01, 0.03, 0.05, 0.07}\), and was found to yield the best tradeoff between stability and performance. Noise injection serves as a regularization method to enhance model generalization to minor deviations in feature values, a common occurrence in real-time wireless traffic. 

This regularization strategy aimed to increase the model's tolerance to real-world variabilities in wireless traffic. Fig. \ref{fig:ROC_AUC} illustrates the ROC curves for both settings. The Area Under the Curve (AUC) improved from 0.93 (without noise) to 0.98 (with noise), indicating significantly improved separability. This is supported by ablation results: the F1-score increased from 0.9497 to 0.9603, and FPR decreased from 4.46\% to 3.49\%. These enhancements confirm that noise injection effectively strengthens model generalization and boundary learning under noisy conditions.

\begin{figure}[h]
\centering

\begin{minipage}[b]{0.48\textwidth}
  \centering
  \includegraphics[width=\textwidth]{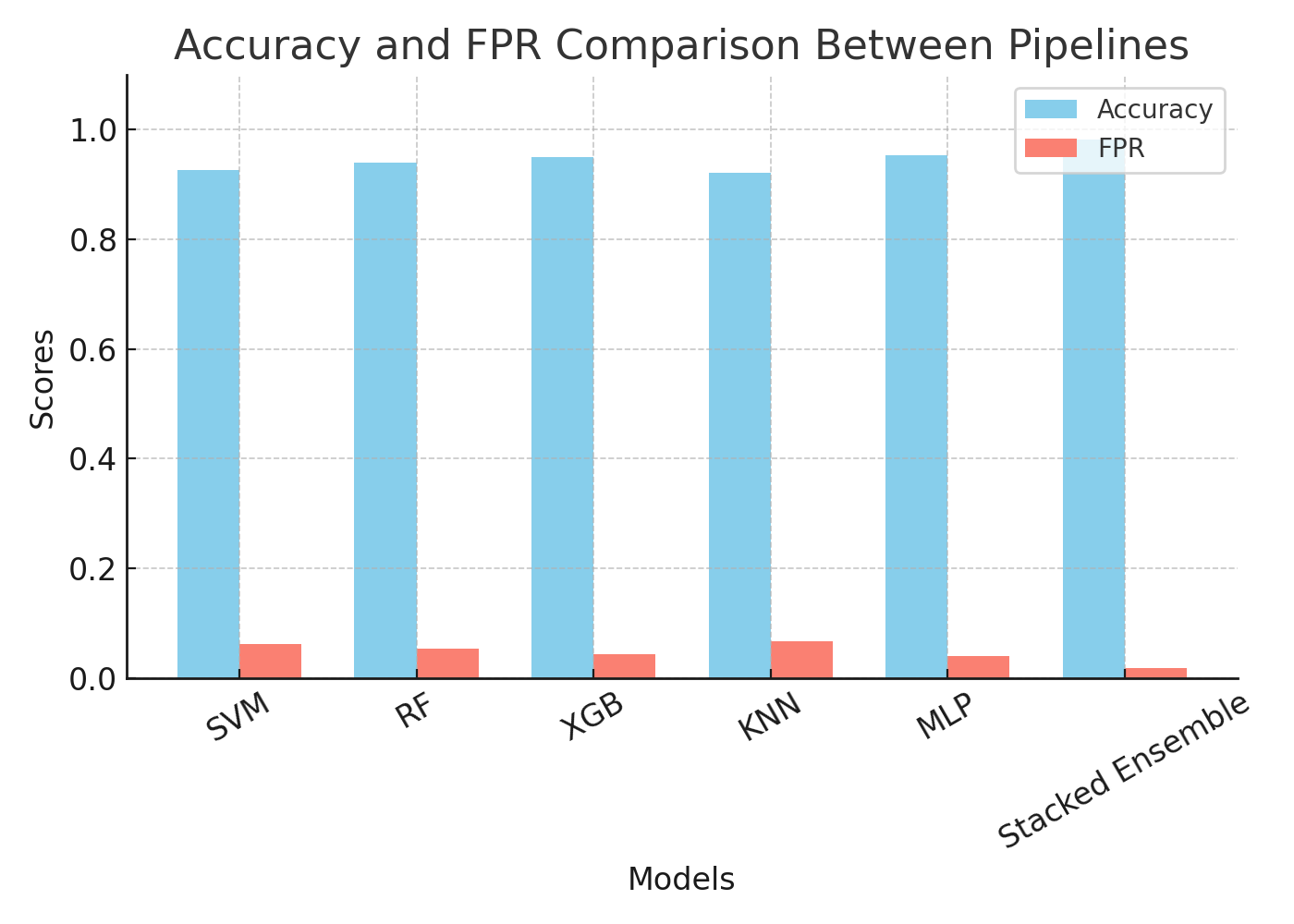}
  \caption{Comparison of Accuracy and FPR between Pipelines}
  \label{fig:comparison}
\end{minipage}
\hfill
\begin{minipage}[b]{0.48\textwidth}
  \centering
  \includegraphics[width=\textwidth]{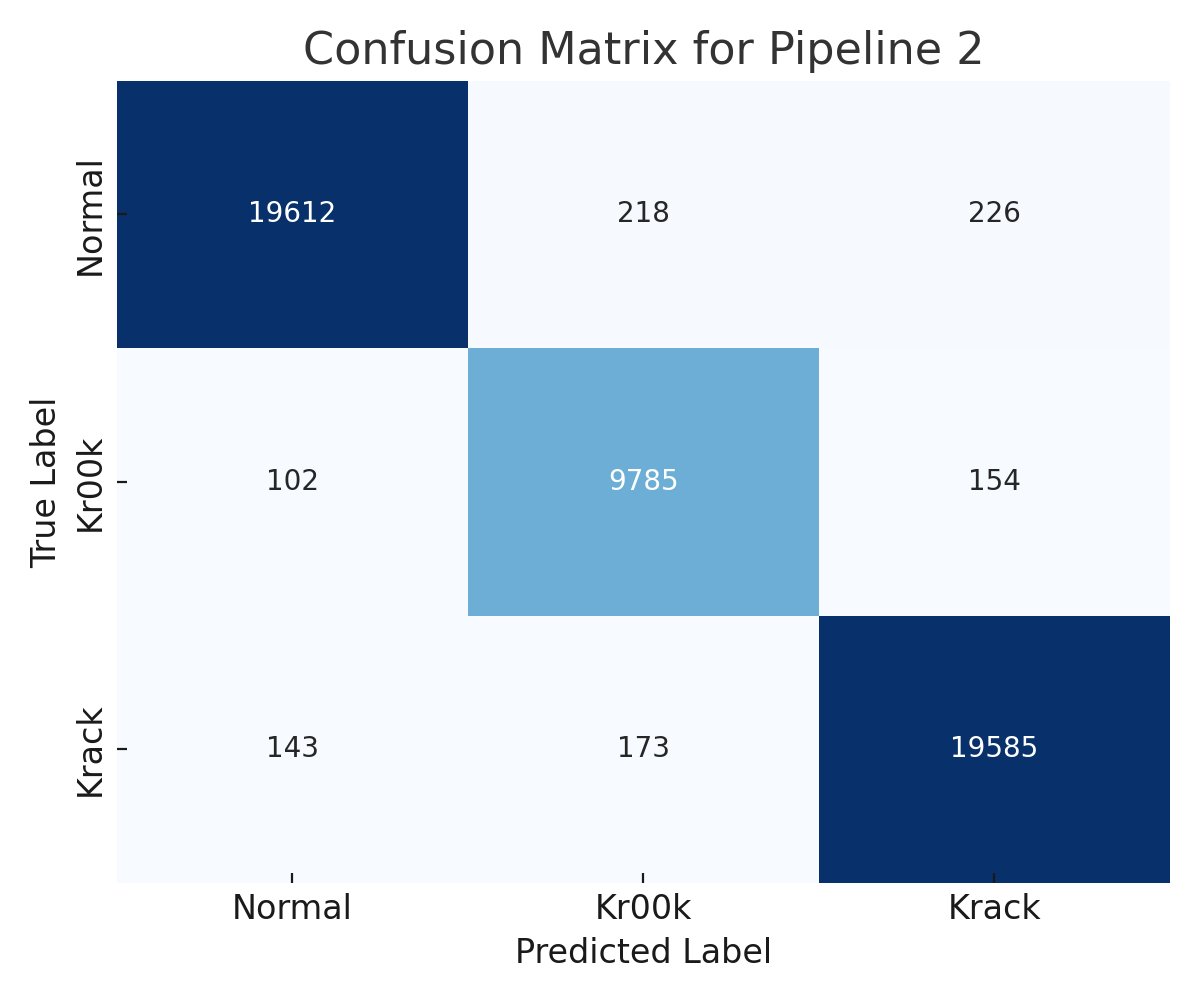}
  \caption{Confusion Matrix for Pipeline 2}
  \label{fig:conf_matrix}
\end{minipage}

\end{figure}

\noindent \textbf{Comparative Insights.} To better understand the gains from ensemble learning, Figure~\ref{fig:comparison} compares Accuracy and FPR across both pipelines. Pipeline 1 models like MLP and XGBoost achieve relatively high accuracy, but all exhibit FPRs between 0.04 to 0.07\%. By contrast, Pipeline 2 maintains accuracy in the 98 to 99\% range and consistently lowers FPRs below 0.02\% for all classes.
This improvement can be attributed to the meta-classifier's ability to learn from the disagreement patterns of the base models, as well as the enhanced data representation obtained via noise injection and PCA. 

Additionally, the confusion matrix in Fig.~\ref{fig:conf_matrix} reveals that most samples are correctly classified, with only a small number of misclassifications across classes.
These results confirm that Pipeline 2 offers balanced performance across all classes, minimizing both false negatives and false positives, a critical property for real-time security systems.

To better contextualize the performance gains achieved by our proposed Pipeline 2, Table~\ref{tab:comparative-methods} highlights key limitations in previous IDS approaches and contrasts them with the improvements introduced in our work. This comparison spans critical aspects such as noise robustness, dimensionality reduction, ensemble learning strategies, and inference efficiency. The methodological enhancements directly align with our design objectives and support the empirical results presented in subsequent sections.

\begin{table}[h]
\centering
\caption{Comparative analysis of limitations in prior methods and improvements introduced by the proposed pipeline.}
\label{tab:comparative-methods}
\begin{tabular}{p{3.5cm}p{5.8cm}p{5.8cm}}
\hline
\textbf{Aspect} & \textbf{Limitations in Previous Approaches} & \textbf{Improvements in Proposed Method} \\
\hline
Noise Robustness & Often absent; models fail under noisy or adversarial settings & Gaussian noise ($\sigma = 0.05$) improves robustness and class separability \\

Dimensionality Reduction & Rarely applied; high-dimensional features increase overfitting & PCA (90\% retained variance) reduces feature noise and boosts stability \\

Ensemble Learning & Applied partially; without effective aggregation strategies & Stacked ensemble of SVM, RF, KNN, MLP, XGBoost \\

Meta-Classifier Strategy & Typically absent; single model reliance increases bias & XGBoost as meta-learner ensures flexible, regularized decision boundary \\

False Positive Rate (FPR) & Often above 0.04\%, unsuitable for critical deployments & Reduced to  FPR below 0.02\% (Table~\ref{tab:comparison_all_metrics}) \\

Generalization Stability & Higher variance under cross-validation & Demonstrated low std-dev across 5-fold CV (see Table~\ref{tab:pipeline2-results}) \\

Multiclass Detection Capability & Limited to binary classification in many cases & 3-class classification (Normal, Kr00k, Krack) with high fidelity \\

Resource Efficiency (Deployability) & Rarely evaluated; few report inference cost for IoT environments & Inference time: $\sim$28\,ms/sample; suitable for IoT edge deployment \\
\hline
\end{tabular}
\end{table}

\subsection*{Ablation Study}

To understand the contribution of each component in our proposed Pipeline 2, we conducted an ablation study with four progressively enhanced configurations:
(i) base learners only,
(ii) noise injection,
(iii) PCA-based dimensionality reduction, and
(iv) full stacked ensemble with probabilistic meta-learning.

Table~\ref{tab:ablation} reports the results averaged over 5 runs using stratified 5-fold cross-validation.
\begin{table}[h]
\centering
\small
\caption{Ablation Study on Pipeline Components}
\label{tab:ablation}
\begin{tabular}{lccccc}
\toprule
\textbf{Configuration} & \textbf{Accuracy} & \textbf{F1-score} & \textbf{Precision} & \textbf{Recall} & \textbf{FPR} \\
\midrule
Base Learners Only        & 0.9498 & 0.9497 & 0.9482 & 0.9513 & 0.0446 \\
+ Noise Injection         & 0.9614 & 0.9603 & 0.9587 & 0.9621 & 0.0349 \\
+ PCA                     & 0.9702 & 0.9691 & 0.9675 & 0.9708 & 0.0273 \\
+ Stacked Ensemble (Full) & \textbf{0.9802 ± (0.0014)} & \textbf{0.9786 ± (0.0015)} & \textbf{0.9774 ± (0.0013)} & \textbf{0.9798 ± (0.0015)} & \textbf{0.0196 ± (0.0009)} \\
\bottomrule
\end{tabular}
\end{table}

The ablation results show that each component progressively contributes to performance improvement. Noise injection improves generalization by introducing variation during training. PCA reduces feature noise and redundancy, further enhancing precision and recall. Finally, the stacked ensemble delivers the most significant performance gains, reducing the FPR by over 56\% compared to the base learners.

\subsection*{Runtime Performance and Deployability}

To evaluate practical deployment feasibility, we profiled runtime performance on a standard mid-tier CPU setup (Intel Core i7, 32GB RAM). The complete training process for our stacked ensemble, including preprocessing, Gaussian noise injection, PCA, five base classifiers, and a meta-classifier, required approximately 20 minutes (1200 seconds). While training time is moderate, this step is conducted offline and does not affect real-time operation. Once trained, the model demonstrates efficient runtime behavior. The average inference time per sample is approximately 28 milliseconds, with a model size of 17MB, making it feasible for use in resource-constrained edge devices such as ARM Cortex-A processors or Raspberry Pi systems.

The use of PCA to reduce feature dimensionality, and the compact 15-dimensional meta-feature vector for the meta-classifier, ensures low computational load during inference. We also monitored CPU usage and observed minimal memory overhead during evaluation. These results support our claim that the proposed IDS is deployable in both edge and cloud environments depending on latency and throughput requirements.

\section*{Conclusion}
\label{sec:Conclusions}

This study explored the effectiveness of machine learning-based IDS for detecting complex wireless attacks, specifically KRACK and Kr00k, in IoT Wi-Fi environments. Exploring the AWID3 data set and a robust preprocessing pipeline, we addressed key challenges such as noise variability, feature redundancy, and class imbalance, which often hinder real-world IDS deployment.

Through extensive experimentation, we demonstrated that our proposed stacked ensemble architecture (Pipeline 2) significantly outperforms individual classifiers on all major performance metrics. By integrating heterogeneous learners via meta-learning and combining this with noise injection and dimensionality reduction through Principal Component Analysis (PCA), the model achieved consistent improvements in accuracy, precision, recall, and particularly in reducing false positive rates (FPR), a critical metric for operational IDS performance. While slight degradations from ideal scores were observed due to stochastic elements like noise augmentation and cross-validation splits, the ensemble maintained superior generalization and robustness. The layered approach effectively leveraged the strengths of individual classifiers while mitigating their weaknesses, resulting in a scalable, high-performance detection system suitable for intelligent networks.

Overall, this work underlines the importance of combining preprocessing strategies (such as noise handling, feature selection, and PCA) with ensemble learning techniques to build reliable and adaptable intrusion detection models. The proposed methodology not only improves detection accuracy but also reduces computational overhead, making it suitable for deployment in resource-constrained IoT environments.

\noindent \textbf{Future Work.} Although we did not conduct GPU-based benchmarking, the model architecture is compatible with GPU-accelerated inference using frameworks like ONNX or joblib-parallel, which could reduce inference time substantially (potentially to <1ms per sample in batch mode). We note this as a direction for future optimization. In addition, future studies should focus on deploying these ensemble pipelines in real-time IDS frameworks, testing their resilience under adversarial conditions, and adapting them to evolving threat landscapes. Incorporating domain adaptation, continual learning, or uncertainty-aware decision-making could further improve robustness in dynamic and heterogeneous IoT network environments. Future work may also evaluate the proposed approach on data sets such as CIC-IDS2018, IoT-23, and UNSW-NB15.

\section*{Data Availability}
\label{sec:DataAvailability}
The data set analysed during the current study is publicly available as part of the AWID3 benchmark data set at: \url{https://icsdweb.aegean.gr/awid/download-data set}.

\bibliography{sample}

\section*{Acknowledgements}

The authors would like to thank anonymous reviewers and the editors of the journal. Your constructive comments have improved the quality of this paper.

\section*{Author contributions statement}

Md Minhazul Islam Munna, Mahbubur Rahman; conceptualization and methodology: Md Minhazul Islam Munna, Mahbubur Rahman, Jaroslav Frnda; data collection: Md Minhazul Islam Munna, Mahbubur Rahman; analysis and interpretation of results: Md Minhazul Islam Munna, Mahbubur Rahman, Jaroslav Frnda, Muhammad Shahid Anwar ; draft manuscript preparation: Md Minhazul Islam Munna, Mahbubur Rahman, Muhammad Shahid Anwar ; review and editing: Md Minhazul Islam Munna, Mahbubur Rahman, Jaroslav Frnda, Muhammad Shahid Anwar , Alpamis Kutlimuratov; project administration: Muhammad Shahid Anwar. All authors reviewed the results and approved the final version of the manuscript.

\section*{List of Abbreviations}
\textbf{AI} : Artificial Intelligence \\
\textbf{ANOVA} : Analysis of Variance \\
\textbf{ARM} : Advanced RISC Machine \\
\textbf{AUC} : Area Under the Curve \\
\textbf{AWID} : Aegean Wireless Intrusion Dataset \\
\textbf{BGOA} : Binary Grasshopper Optimization Algorithm \\
\textbf{BILSTM} : Bidirectional Long Short-Term Memory \\
\textbf{CIA} : Confidentiality, Integrity, Availability \\
\textbf{CIC} : Canadian Institute for Cybersecurity \\
\textbf{CNN} : Convolutional Neural Network \\
\textbf{CPS} : Cyber-Physical System \\
\textbf{CPU} : Central Processing Unit \\
\textbf{CSA} : Cloud Security Alliance \\
\textbf{CSE} : Computer Science and Engineering \\
\textbf{CSV} : Comma Separated Values \\
\textbf{CV} : Cross Validation / Computer Vision (context-dependent) \\
\textbf{CWFLAM} : Class Weighted Focal Loss Adaptive Mechanism \\
\textbf{DL} : Deep Learning \\
\textbf{DNN} : Deep Neural Network \\
\textbf{EAP} : Extensible Authentication Protocol \\
\textbf{EDA} : Exploratory Data Analysis \\
\textbf{ESET} : ESET Security Software \\
\textbf{FFA} : Firefly Algorithm \\
\textbf{FPR} : False Positive Rate \\
\textbf{GBT} : Gradient Boosted Trees \\
\textbf{GPU} : Graphics Processing Unit \\
\textbf{GRU} : Gated Recurrent Unit \\
\textbf{GWO} : Grey Wolf Optimizer \\
\textbf{IDS} : Intrusion Detection System \\
\textbf{IEEE} : Institute of Electrical and Electronics Engineers \\
\textbf{IP} : Internet Protocol \\
\textbf{IQR} : Interquartile Range \\
\textbf{IRF} : Impulse Response Function \\
\textbf{KDD} : Knowledge Discovery in Databases \\
\textbf{KNN} : K-Nearest Neighbors \\
\textbf{KRACK} : Key Reinstallation Attack \\
\textbf{LSTM} : Long Short-Term Memory \\
\textbf{MAC} : Media Access Control \\
\textbf{MCA} : Multiple Correspondence Analysis \\
\textbf{ML} : Machine Learning \\
\textbf{MLP} : Multilayer Perceptron \\
\textbf{NN} : Neural Network \\
\textbf{NSL} : NSL-KDD Dataset \\
\textbf{ONNX} : Open Neural Network Exchange \\
\textbf{OSI} : Open Systems Interconnection \\
\textbf{PCA} : Principal Component Analysis \\
\textbf{PPDNN} : Privacy Preserving Deep Neural Network \\
\textbf{RAM} : Random Access Memory \\
\textbf{RF} : Random Forest \\
\textbf{ROC} : Receiver Operating Characteristic \\
\textbf{SCADA} : Supervisory Control and Data Acquisition \\
\textbf{SHAP} : SHapley Additive exPlanations \\
\textbf{SM} : Security Mechanism (or Social Media, context-dependent) \\
\textbf{SMOTE} : Synthetic Minority Over-sampling Technique \\
\textbf{SQL} : Structured Query Language \\
\textbf{SSDP} : Simple Service Discovery Protocol \\
\textbf{SSH} : Secure Shell \\
\textbf{SVM} : Support Vector Machine \\
\textbf{TL} : Transfer Learning \\

\end{document}